\documentstyle[twoside,fleqn,espcrc2,epsf]{article}

\newcommand{\AmS}{{\protect\the\textfont2
  A\kern-.1667em\lower.5ex\hbox{M}\kern-.125emS}}

\begin{document}

\title{Potential Nonrelativistic QCD and 
Heavy Quarkonium Spectrum in  Next-to-Next-to-Next-to-Leading
Order
\thanks{Talk presented at  {\em RADCOR/LL 2002} conference,
Kloster Banz, Germany, September 8-13, 2002.}
}

\author{Alexander A. Penin
\address{II.\ Institut f\"ur Theoretische Physik, Universit\"at Hamburg,\\ 
Luruper Chaussee 149, D-22761 Hamburg, Germany}\thanks{Permanent address:
Institute for Nuclear Research,  60th October Anniversary Pr., 
7a, Moscow 117312, Russia}}

\begin{abstract}
In this talk I  review the  
calculation of the third  
order corrections to the heavy quarkonium spectrum in the 
nonrelativistic effective theory framework
and its  application to the phenomenology
of top quark threshold production.  
\end{abstract}

\maketitle

\section{INTRODUCTION}

The theoretical study of the nonrelativistic  
heavy quark-antiquark system \cite{AppPol}  and its application
to bottomonium \cite{NSVZ} and toponium   \cite{FadKho}
physics is of special interest because it relies 
entirely on the  first principles of QCD. 
The system in principle allows for a perturbative treatment
with the nonperturbative effects  being well under control
and with no crucial model dependence. This  makes the heavy
quarkonium to be  an ideal
place to determine the fundamental QCD parameters
such as the heavy quark mass $m_q$ and the strong coupling
constant $\alpha_s$.  An essential progress 
has been made in the theoretical 
investigation of the nonrelativistic heavy quark dynamics
based on  the effective theory approach  \cite{CasLep}.
The analytical results for the main parameters of
the nonrelativistic heavy quark-antiquark system
are now available up to next-to-next-to-leading order (NNLO) 
(see \cite{Pen} for a brief review).
The NNLO corrections have turned out to be so sizeable that it appears to be
indispensable also to gain control over the next-to-next-to-next-to-leading
order (N$^3$LO) both in regard of phenomenological applications and in order 
to understand the structure and the peculiarities of the nonrelativistic
expansion. The calculation of the heavy-quarkonium spectrum 
to   ${\cal O}(\alpha_s^3)$ \cite{KPSS2,PenSte} was 
the first breakthrough in the  N$^3$LO analysis. 
Recently, the  ${\cal O}(\alpha_s^3\ln{\alpha_s})$ corrections
to the  heavy-quarkonium production and annihilation rates
have been also obtained \cite{KPSS3}. The subject of this talk is  restricted
mainly to the analysis of the  spectrum. 

In the next section we briefly introduce the basic ingredients of the
nonrelativistic effective-theory formalism.
In Sect.~3 the structure of the  third order corrections
to the heavy quarkonium ground state energy is outlined and the final result
is presented. In the last section this result is applied for the 
phenomenological analysis of the top-antitop pair production
near the threshold.

\section{EFFECTIVE THEORY OF NONRELATIVISTIC HEAVY QUARKS}

The nonrelativistic behavior of the heavy-quark-antiquark pair is governed by
a complicated multiscale dynamics.
In the nonrelativistic regime, where the heavy-quark velocity $v$ is of the
order of the strong-coupling constant $\alpha_s$, the Coulomb effects are
crucial and have to be taken into account to all orders in $\alpha_s$.
This makes the use of the effective theory mandatory.
The effective-theory approach allows us to separate the scales and to
implement the expansion in $v$ at the level of the Lagrangian.
Let us recall that the dynamics of a nonrelativistic quark-antiquark pair is
characterized by four different {\it regions} and the corresponding modes
\cite{BenSmi}: (i)
the hard region (the energy and three-momentum scale like $m_q$); 
(ii) the soft region (the energy and three-momentum scale like $m_qv$);
(iii) the potential region (the energy scales like $m_qv^2$, while the
three-momentum scales like $m_qv$); and
(iv) the ultrasoft region (the energy and three-momentum scale like $m_qv^2$). 
Nonrelativistic QCD (NRQCD) \cite{CasLep} is obtained by integrating out
the hard modes.
Subsequently integrating out the soft modes and the potential gluons results
in the effective theory of potential NRQCD (pNRQCD) \cite{PinSot1}, which
contains potential heavy quarks and ultrasoft gluons, ghosts, and light quarks
as active particles.
The effect of the modes that have been integrated out is two-fold:
higher-dimensional operators appear in the effective Hamiltonian,
corresponding to an expansion in $v$, and the Wilson coefficients of the
operators in the effective Hamiltonian acquire  corrections, which are series
in $\alpha_s$.

The theory of pNRQCD is relevant for the description of the heavy-quarkonium
system.
Let us recall its basic ingredients.
In pNRQCD, the (self)interactions between ultrasoft particles are described by
the standard QCD Lagrangian.
The interactions of the ultrasoft gluons with the heavy-quark-antiquark pair
are ordered in $v$ by the multipole expansion.
For the N$^3$LO analysis, only the leading-order (LO) emission and absorption
of ultrasoft gluons have to be considered.
They are described by the chromoelectric dipole interaction.
The propagation of the quark-antiquark pair in the colour-singlet  and
colour-octet  states is described by the nonrelativistic Green function
of the effective Schr{\"o}dinger equation.
The LO approximation for the Green function is given by the Coulomb solution,
which sums up terms singular at threshold and describes the leading binding
effects.
The corrections  due to higher-order terms in the
effective  Hamiltonian can be found in Rayleigh-Schr\"odinger time-independent
perturbation theory as in standard quantum mechanics.

Let us now turn to the problem of perturbative calculations in the effective
theory.
Both NRQCD and pNRQCD have specific Feynman rules, which can be used for a
systematic perturbative expansion.
However, this is complicated because the expansion of the Lagrangian
corresponds to a particular subspace of the total phase space.
Thus, in a perturbative calculation within the effective theory, one has to
formally impose some restrictions on the allowed values of the virtual
momenta.
Explicitly separating the phase space introduces additional scales to the
problem, such as momentum cutoffs, and makes the approach considerably less
transparent.
A much more efficient and elegant method 
is based on the expansion by regions
\cite{BenSmi}, which is a systematic method to expand Feynman diagrams
in any limit of momenta and masses.
It consists of the following steps:
(i) consider various regions of a loop four-momentum $k$ and expand, in every
region, the integrand in Taylor series with respect to the parameters that
are considered to be small there;
(ii) integrate the expanded integrand over the whole integration domain 
of the loop momenta; and (iii) put to zero any scaleless integral
In step~(ii), dimensional regularization, with $d=4-2\epsilon$ space-time
dimensions, is used to handle the divergences.
In the case of the threshold expansion in $v$, one has to deal with the four
regions and their scaling rules listed above.
In principle, the threshold expansion has to be applied to the Feynman
diagrams of full QCD.
However, after integrating out the hard modes, which corresponds to
calculating the hard-region contributions in the threshold expansion, it is
possible to apply step (i) to the diagrams constructed from the NRQCD and
pNRQCD Feynman rules.
Equivalently, the Lagrangian of the effective theory can be employed for a
perturbative calculation without explicit restrictions on the virtual momenta
if dimensional regularization is used and the formal expressions derived from
the Feynman rules of the effective theory are understood in the sense of the
threshold expansion.
In this way, one arrives at a formulation of effective theory with two crucial
virtues: the absence of additional regulator scales and the automatic matching
of the contributions from different scales   
\cite{KPSS2,KPSS1} (see also\cite{PinSot2}).
The second property implies that the contributions of different modes, as
computed in the effective theory, can be simply added up to get the full
result.

\section{HEAVY QUARKONIUM SPECTRUM AT ${\cal O}(\alpha_s^5m_q)$}
The analysis of the heavy quarkonium spectrum at 
${\cal O}(\alpha_s^5m_q)$  {\em i.e.} 
the third order corrections to the Coulomb 
approximation involves two basic ingredients:
the effective Hamiltonian and the 
retardation effect associated with the emission and absorption of
dynamical ultrasoft gluons.
The general form of the Hamiltonian valid up to 
N$^3$LO  is given in~\cite{KPSS2}.
A detailed discussion of the effects 
resulting from the chromoelectric dipole interaction 
of the heavy quark-antiquark  pair to the ultrasoft gluons 
relevant for perturbative bound-state calculations at N$^3$LO
can be found in \cite{KPSS2,KniPen1}.
 
For vanishing angular momentum we can write the 
energy level of the principal quantum number $n$ as
\begin{equation}
E_n^{\rm p.t.}=E^C_n+\delta E^{(1)}_n+\delta E^{(2)}_n+\delta
E^{(3)}_n
+\ldots\,,
\label{enseries}
\end{equation}
where the leading order Coulomb energy is 
$E^C_n=-{C_F^2\alpha_s^2m_q/4n^2}$.
$\delta E^{(k)}_n$ stands for corrections of order $\alpha_s^k$.
The first and second order corrections can be found in  
\cite{PinYnd,MelYel,KPP} for arbitrary $n$.   

The ${\cal O}\left(\alpha_s^3\right)$ corrections 
to the energy levels arise from several sources:
(i) matrix elements of the N$^3$LO operators of the effective
Hamiltonian between Cou\-lomb wave functions; 
(ii) higher iterations of the NLO and NNLO operators of the effective
Hamiltonian in time-independent perturbation theory; 
(iii) matrix elements of the N$^3$LO instantaneous operators generated
by the emission and absorption of ultrasoft gluons; and
(iv) the retarded ultrasoft contribution.

The contributions to $E_n^{(3)}$ for vanishing 
$\beta$-function were computed for general $n$ in \cite{KPSS2}. 
The corresponding correction to the ground state energy
gets contributions from all four sources and reads
\begin{eqnarray}
\lefteqn{\left.\delta E^{(3)}_1\right|_{\beta(\alpha_s)=0}
=-E^C_1{\alpha_s^3\over\pi}\left\{
-{a_1a_2+a_3\over32\pi^2}\right.}\nonumber\\
\lefteqn{\left.
+\left[-{C_AC_F\over2}+\left(-\frac{19}{16}
+{S(S+1)\over2}\right)C_F^2\right]{a_1}\right.}
\nonumber\\
\lefteqn{+\left[-{1\over 36}+\frac{\ln2}{6}+\frac{L_{\alpha_s}}{6}\right]
{C_A^3}+\left[-{49\over36}\right.
}
\nonumber\\
\lefteqn{\left.+{4\over3}\left(\ln2+L_{\alpha_s}\right)
\right]{C_A^2C_F}+\left[-{5\over 72}+{10\over 3}\ln2
\right.}
\nonumber\\
\lefteqn{\left.+{37\over6}L_{\alpha_s}+
\left({85\over 54}
-{7\over 6}L_{\alpha_s}\right)S(S+1)\right]
{C_AC_F^2}}
\nonumber\\
\lefteqn{+\left[{50\over9}+{8\over3}\ln2
+3L_{\alpha_s}-{S(S+1)\over3}\right]
{C_F^3}}
\nonumber\\
\lefteqn{+\left[-{32\over15}+2\ln2+(1-\ln2)S(S+1)\right]{C_F^2T_F}}
\nonumber\\
\lefteqn{+{49C_AC_FT_Fn_l\over36}
+\left[{11\over18}-{10\over27}S(S+1)\right]{C_F^2T_Fn_l}}
\nonumber\\
\lefteqn{\left.+{2\over3}C_F^3L^E_1\right\}\,,}
\label{zerobeta}
\end{eqnarray}
where the color group invariants are
$C_F=4/3$, $C_A=3$, $T_F=1/2$, 
$n_l$ is the number of light-quark flavors,
$S$ is  the spin quantum number and 
$L_{\alpha_s}=-\ln(C_F\alpha_s)$.
The numerical value of the 
QCD Bethe logarithm $L^E_1=-81.5379$ can be found in~\cite{KPSS2,KniPen1}.
$\overline{\rm MS}$ scheme for the renormalization of $\alpha_s(\mu)$ 
is implied. 
The logarithmic $\ln(\alpha_s)$ part of Eq.~(\ref{zerobeta})
has been derived first in  \cite{BPSV2,KniPen2}.
The coefficient
$a_i$ parameterizes the $i$-loop correction to the Coulomb potential
($a_1=31C_A/9-20T_Fn_l/9,\ldots$ \cite{Sch}).
At present, only Pad\'e estimates of the three-loop $\overline{\rm MS}$
coefficient $a_3$ are available \cite{ChiEli}. For the bottom and top quark
case they read 
\begin{eqnarray}
{a_3\over 4^3}&=&
\cases{
\displaystyle
\displaystyle
98&if $n_l=4$\,,\cr
\displaystyle
60&if $n_l=5$\,.\cr}
\label{a3}
\end{eqnarray}
The third order corrections to the ground state energy
proportional to the $\beta$-function 
only get contributions from (i) and (ii)
and read
\begin{eqnarray}
\lefteqn{\left.\delta E^{(3)}_1\right|_{\beta(\alpha_s)}
=E^C_1\left({\alpha_s\over\pi}\right)^3\bigg\{
{32\beta_0^3}L^3_\mu+\left[{40}\beta_0^3
\right.}
\nonumber \\
\lefteqn{\left.+{12}a_1\beta_0^2+
{28}\beta_1\beta_0\right]L^2_\mu
+\left[\left({16\pi^2\over 3}+64\zeta(3)\right)\beta_0^3\right.}
\nonumber \\
\lefteqn{
+{10}a_1\beta_0^2
+\left({40}\beta_1+{a_1^2\over 2}+{a_2}+8\pi^2C_AC_F\right.}
\nonumber \\
\lefteqn{
\left.\left.
+\left({21\pi^2\over 2}-{16\pi^2\over 3}S(S+1)\right)C_F^2\right)\beta_0
+{3}a_1\beta_1\right.}
\nonumber \\
\lefteqn{+{4\beta_2}\bigg]L_\mu
+\left(-{8}+{4\pi^2}+{2\pi^4\over45}+64\zeta(3)\right.}
\nonumber \\
\lefteqn{
-{8\pi^2}\zeta(3)+{96}\zeta(5)\bigg)\beta_0^3
+\left({2\pi^2\over 3}+{8\zeta(3)}\right)a_1\beta_0^2}
\nonumber \\
\lefteqn{
+\left(
\left({8}+{7\pi^2\over 3}
+{16\zeta(3)}\right)\beta_1
-{a_1^2\over 8}+{3\over 4}a_2+\bigg({6\pi^2}\right.}
\nonumber \\
\lefteqn{
\left.-{2\pi^4\over 3}\right)C_AC_F
+\left(8\pi^2-{4\pi^4\over 3}+\left(-{4\pi^2\over 3}
+{4\pi^4\over 9}\right)\right.}
\nonumber \\
\lefteqn{
\left.\left.\left.
\times S(S+1)\right)
C_F^2\right)\beta_0
+{2a_1\beta_1}+{4\beta_2}
\right\}\,,}
\label{beta}
\end{eqnarray}
where $L_\mu=\ln(\mu/(C_F\alpha_sm_q))$,
$\zeta(z)$ is Riemann's $\zeta$-function and 
$\beta_i$ is the $(i+1)$-loop coefficient of the QCD $\beta$ function
($\beta_0=11C_A/12-T_Fn_l/3,\ldots$).
The terms proportional to $\beta_0^3$ in Eqs.~(\ref{beta}) 
have been computed first  in  \cite{KiySum}.

The total result for the third order correction 
to the ground state energy 
is given by the sum of Eqs.~(\ref{zerobeta}) and (\ref{beta})
Adopting the choice $\mu_s=C_F\alpha_s(\mu_s) m_q$ one obtains 
for the bottom and top system (for S=1) in numerical form
\begin{eqnarray}
\lefteqn{ \delta E^{(3)}_1= \alpha_s^3E_1^C\left[
  \left(
  \begin{array}{c} 
    70.590|_{n_l=4}\\
    56.732|_{n_l=5} 
  \end{array} 
  \right) + 15.297 \right.}
 \nonumber\\
\lefteqn{ \left. \times \ln(\alpha_s) + 0.001\,a_3 
  + \left(
  \begin{array}{c} 
    34.229|_{n_l=4}\\
    26.654|_{n_l=5} 
  \end{array} 
  \right)\bigg|_{\beta_0^3}
  \right]\,,}
  \label{eq:dele3num}
\end{eqnarray}
where we have separated the contributions arising from $a_3$ and
$\beta_0^3$.
The only unknown ingredient in our result for $\delta E^{(3)}_1$
is the three-loop $\overline{\rm MS}$ coefficient $a_3$ 
of the corrections to the static potential entering Eq.~(\ref{zerobeta}).
Up to now there are only estimates based on Pad\'e approximation which
we will use in our analysis.
However, our final result only changes marginally
even for a rather large deviations of $a_3$ from its  Pad\'e estimate.

\section{PHENOMENOLOGY OF  $t\bar t$ THRESHOLD PRODUCTION}
\begin{figure}[t]
   \vspace{-1.8cm}
   \epsfysize=9.5cm
   \epsfxsize=6.7cm
   \centerline{\epsffile{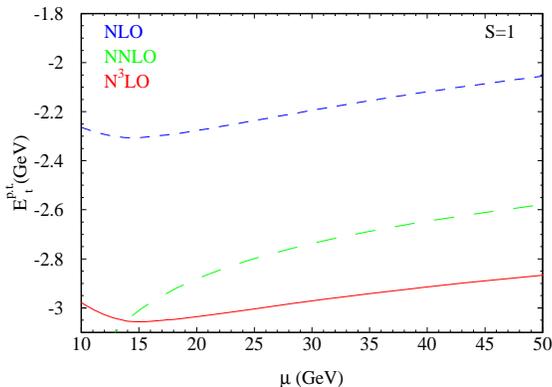}}
   \vspace{-3.0cm}
\caption[dummy]{\label{fig1}\small 
Ground state energy $E^{\rm p.t.}_1$ of the 
$t\bar{t}$ bound state in the zero-width approximation 
as a function of the renormalization scale $\mu$
for and $S=1$. 
}
\vspace{-0.8cm}
\end{figure}

Let us  focus on the top quark threshold production.
(the bottom quark case  is considered  in \cite{Ste}).  
The corresponding experimental data are not yet available  
and our goal is to present
a formula which can be directly used for the top quark mass 
determination from the characteristics of the 
cross section of $t\bar t$ production near threshold \cite{PenSte}.
One has to distinguish the production
in  $e^+e^-$ annihilation
where the final state quark-antiquark pair is produced
with $S=1$ and in (unpolarized) $\gamma\gamma$
collisions where the dominant contribution is given by the
$S$-wave zero spin final state. 
The nonperturbative effects in the case of the top quark are
negligible, however, the effect of the
top quark width has to be taken into account properly~\cite{FadKho}
as the relatively large width smears out the Coulomb-like resonances
below the threshold. 
The NNLO analysis of the cross section~\cite{hoaetal}
shows that only the ground state pole results in the well-pronounced 
resonance.
The higher poles and continuum, however, affect the position of the 
resonance peak  and  move it to higher energy.
As a consequence, the  resonance energy can be written
in the form
\begin{eqnarray}
  E_{\rm res}&=&2m_t+E_1^{\rm p.t.}+\delta^{\Gamma_t}E_{\rm res}\,.
  \label{ttcorr}
\end{eqnarray}
To compute the shift of the peak position 
due to the nonzero top quark width $\delta^{\Gamma_t} E_{\rm res}$
we use the result of \cite{PenPiv2} for the $e^+e^-\to t\bar t$
and $\gamma\gamma\to t\bar t$ cross sections 
at NNLO. Numerically for both processes
\begin{equation}
  \delta^{\Gamma_t} E_{\rm res}=100\pm 10~{\rm MeV}\,.
  \label{engam}
\end{equation}
This value is rather stable with respect to the variation of 
all input parameters of our analysis.
To evaluate  the perturbative contribution we use
the result of the previous section up to N$^3$LO. 
In Fig.~\ref{fig1} the perturbative
ground state energy  in the next-to-leading (NLO), NNLO and N$^3$LO is 
plotted as a function of the renormalization scale 
of the strong coupling constant
for  $S=1$.
One can see, that the N$^3$LO result shows a much weaker 
dependence on $\mu$ than the NNLO one. 
Moreover at the scale $\mu\approx 15$~GeV 
which is  close to the physically motivated scale $\mu_s$,
the N$^3$LO correction vanishes and furthermore 
becomes independent of $\mu$, {\it i.e.} the N$^3$LO curve shows a local
minimum. This suggests
the convergence of the  series for 
the ground state energy {\em in the pole mass scheme}.
Collecting all the contributions  we obtain a universal 
relation between the resonance energy in $t\bar t$ threshold 
production in $e^+e^-$ annihilation or $\gamma\gamma$ collisions
(the result shows very weak spin dependence)
and the top quark pole mass  
\begin{eqnarray}
E_{\rm res}&=& \bigg(1.9833 
+ 0.007\,\frac{m_t-174.3~\mbox{GeV}}{174.3~\mbox{GeV}}
\nonumber\\
&&\pm 0.0009\bigg)\times m_t\,,
  \label{ttcorrnum1}
\end{eqnarray}
where the weak nonlinear dependence of
Eq.~(\ref{ttcorr}) on $m_t$
is taken into account in the second term on the right-hand side of
Eq.~(\ref{ttcorrnum1}).
The central value is computed for $m_t=174.3$~GeV.
The analysis of the theoretical uncertainty in 
Eq.~(\ref{ttcorrnum1}) can be found in \cite{PenSte}.
Due to the very nice behavior of the perturbative expansion
for the ground state energy we do not expect large higher 
order corrections to our result.
As a consequence the top quark pole mass
can be extracted from the future experimental data 
on $t\bar t$ threshold production with the theoretical uncertainty 
of $\pm 80$~MeV.
Thus  we find  the N$^3$LO corrections 
to stabilize the behavior of the perturbative series 
for the ground state energy and to be  mandatory for a high 
precision theoretical analysis.

\vspace{3mm}
\noindent
{\bf Acknowledgments}\\[1mm]
This work was supported in part by DFG Grant No.\ KN~365/1-1 and BMBF Grant
No.\ 05~HT1GUA/4.


\begin{thebibliography}{99}


\bibitem{AppPol}
T. Appelquist and H.D. Politzer,
Phys.\ Rev.\ Lett.\ { 34} (1975) 43.

\bibitem{NSVZ}
V.A. Novikov, {\it et al.},
Phys. Rep. C 41 (1978) 1.

\bibitem{FadKho}
V.S. Fadin and V.A. Khoze,
Pis'ma Zh.\ Eksp.\ Teor.\ Fiz.\ { 46} (1987) 417.

\bibitem{CasLep}
W.E. Caswell and G.P. Lepage,
Phys.\ Lett.\ B { 167} (1986) 437.

\bibitem{Pen}
A.A. Penin,
Nucl.\ Phys.\ Proc.\ Suppl.\ { 96} (2001) 418.

\bibitem{KPSS2} B.A. Kniehl, A.A. Penin, V.A. Smirnov, and M. Steinhauser, 
Nucl.\ Phys.\ { B 635}, (2002)  357.

\bibitem{PenSte} A.A. Penin and M. Steinhauser,
Phys.\ Lett.\ B {538},  (2002) 335.

\bibitem{KPSS3} B.A. Kniehl, A.A. Penin, V.A. Smirnov, and M. Steinhauser, 
Report DESY/02-134, hep-ph/0210161.

\bibitem{BenSmi}
M. Beneke and V.A. Smirnov,
Nucl.\ Phys.\ B { 522} (1998) 321;
V.A. Smirnov, {\it
Applied Asymptotic Expansions in Momenta and Masses}
(Springer Tracts in Modern Physics Vol. 177), November 2001.


\bibitem{PinSot1} A. Pineda and J. Soto,
Nucl.\ Phys.\ B (Proc.\ Suppl.) {64},  (1998) 428.


\bibitem{KPSS1} B.A. Kniehl, A.A. Penin, V.A. Smirnov, and M. Steinhauser, 
Phys.\ Rev.\ D {65},  (2002) 091503.

\bibitem{PinSot2} A. Pineda and J. Soto,
Phys.\ Lett.\ B {420},  (1998) 391;
Phys.\ Rev.\ D {59},  (1999) 016005;
A. Czarnecki, K. Melnikov, and A. Yelkhovsky,
Phys.\ Rev.\ A {59},  (1999) 4316;
M. Beneke, A. Signer, and V.A. Smirnov,
Phys.\ Lett.\ B {454} (1999) 137.
B.A. Kniehl and A.A. Penin,
Phys.\ Rev.\ Lett.\ { 85} (2000) 1210, 
{\it ibid.}\ { 85} (2000) 3065 (E); 
{\it ibid.}\ { 85} (2000) 5094.


\bibitem{KniPen1}
B.A. Kniehl and A.A. Penin,
Nucl.\ Phys.\ B { 563} (1999) 200.


\bibitem{PinYnd}
A. Pineda and F.J. Yndurain,
Phys.\ Rev.\ D { 58} (1998) 094022.


\bibitem{MelYel} K. Melnikov and A. Yelkhovsky,
Phys.\ Rev.\ D {59}, 114009 (1999).


\bibitem{KPP} J.H. K\"uhn, A.A. Penin, and A.A. Pivovarov,
Nucl.\ Phys.\ { B534},  (1998) 356;
A.A. Penin and A.A. Pivovarov,
Phys.\ Lett.\ B {435},  (1998) 413;
Nucl.\ Phys.\ { B 549},  (1999) 217; {\it ibid.}\ { B 550}, (1999) 375.

\bibitem{BPSV2}
N. Brambilla, A. Pineda, J. Soto and A. Vairo,
Phys.\ Lett.\ B { 470} (1999) 215.

\bibitem{KniPen2}
B.A. Kniehl and A.A. Penin,
Nucl.\ Phys.\ B { 577} (2000) 197.

\bibitem{Sch}
Y. Schr\"oder,
Phys.\ Lett.\ B { 447} (1999) 321 and references cited therein.


\bibitem{ChiEli}
F.A. Chishtie and V. Elias,
Phys.\ Lett.\ B { 521} (2001) 434.


\bibitem{KiySum} Y. Kiyo and  Y. Sumino,
Phys.\ Lett.\ B 496 (2000) 83. 


\bibitem{Ste} M. Steinhauser, these Proceedings.

\bibitem{hoaetal}
A. H. Hoang
{\it et al.},
Eur.\ Phys.\ J.direct C { 3} (2000) 1.

\bibitem{PenPiv2}
A.A. Penin and A.A. Pivovarov,
Yad.\ Fiz.\ { 64} (2001) 323.



\end{thebibliography}
\end{document}